\documentclass[conference]{IEEEtran}
\IEEEoverridecommandlockouts

\usepackage{cite}
\usepackage{amsmath,amssymb,amsfonts}
\usepackage{graphicx}
\usepackage{booktabs}
\usepackage{textcomp}
\usepackage{xcolor}

\begin{document}

\title{Octree-Context Gaussian Splatting with Multi-Order Propagation for Channel Knowledge Maps}

\author{\IEEEauthorblockN{Jinghan Zhang\IEEEauthorrefmark{1}\IEEEauthorrefmark{2},
Xitao Gong\IEEEauthorrefmark{1},
Qi Wang\IEEEauthorrefmark{1},
Richard A. Stirling-Gallacher\IEEEauthorrefmark{1},
and Giuseppe Caire\IEEEauthorrefmark{2}}
\IEEEauthorblockA{\IEEEauthorrefmark{1}Huawei Heisenberg Research Center, Munich, Germany}
\IEEEauthorblockA{\IEEEauthorrefmark{2}Department of Electrical Engineering and Computer Science, Technical University of Berlin, Berlin, Germany}}

\maketitle

\begin{abstract}
Channel knowledge maps (CKMs) can support environment-aware wireless communications
by providing site-specific channel knowledge as a function of the transmitter position, the receiver position or both.
Existing Gaussian splatting CKM methods can reduce the high training and inference costs of implicit neural methods
yet represent all multipath interactions by one wireless radiance field.
We propose octree-context Gaussian splatting (OctCGS), which anchors Gaussian primitives on an octree as virtual scattering nodes
and decomposes multi-bounce propagation into orders by the number of node-level interactions along each path.
Shared tree attention gives each node local context from neighboring octree cells and room-wide context from ancestor cells at a cost linear in the number of cells,
and the order-wise scattering of each node is predicted from this context.
The order-wise contributions of the power transported between the nodes are summed in power with a line-of-sight (LoS) term
to give the channel gain and the spatial spectrum.
OctCGS raises the spatial-spectrum structural similarity (SSIM) by 0.13 to 0.23 and lowers the mean absolute error (MAE) of the channel gain by 0.9 to 3.8 dB
on five indoor ray-tracing scenes, relative to the best baseline per scene and metric.
\end{abstract}


\section{Introduction}
\label{sec:intro}

Beamforming, handover and resource allocation benefit from reliable channel knowledge,
but the pilot overhead of estimating the channels grows with the numbers of antennas and links.
Channel knowledge maps (CKMs) index site-specific channel knowledge by the transmitter (Tx) position, the receiver (Rx) position or both,
and can thereby reduce this overhead~\cite{zeng2021ckm}.
The available channel samples cover only a small fraction of the input space of a six-dimensional (6D) CKM,
which takes both the Tx position and the Rx position as input~\cite{biwgs}.
Conventional radio map estimation interpolates between samples under spatial stationarity~\cite{romero2022radiomap,CKM_SURVEY},
which the channel can violate wherever walls and objects block or redirect propagation paths.
Accurate 6D CKMs must therefore represent the influence of the environment on the channel at unsampled positions.

The influence of the environment can be derived from a three-dimensional site model or learned from channel samples~\cite{CKM_tutorial}.
Ray tracing resolves the reflections, diffractions and scattering events along each path computed from the site model~\cite{sionna}.
Its accuracy depends on the geometry and the electromagnetic material parameters of this model,
and its computational cost increases with the number of reflecting surfaces in the scene~\cite{RT_survey}.

Implicit neural methods such as NeRF$^2$~\cite{nerf2} and NeWRF~\cite{newrf} can model continuous channel variation from channel samples at known Tx and Rx positions
but the training of NeRF$^2$ is computationally expensive~\cite{CKM_SURVEY}.
Gaussian splatting replaces dense network queries along each ray with explicit Gaussian primitives,
which are projected toward the Rx and blended along each arrival direction.
The methods in~\cite{wrfgsplus,yang2025gsrf,rf3dgs,swiftwrf,bisplatwrf}
apply this representation to wireless radiance fields for one Tx position or one Rx position,
whereas BiWGS~\cite{biwgs} and RxGS~\cite{rxgs} allow both positions to vary.

\begin{figure}[!t]
\centering
\includegraphics[width=\columnwidth]{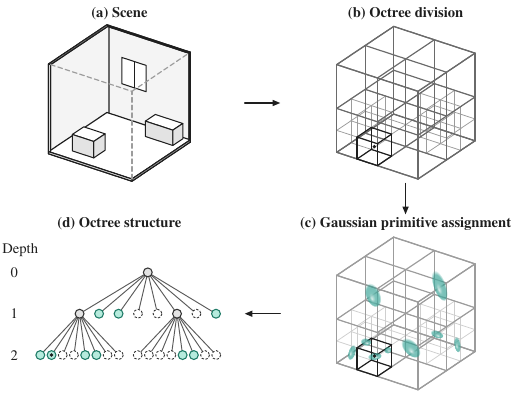}
\caption{Octree division of a room (b), assignment of one Gaussian primitive to each occupied leaf (c) and the resulting octree structure (d).
The dot marks the same leaf in (b)--(d).}
\label{fig:octree}
\end{figure}

\begin{figure*}[!t]
\centering
\includegraphics[width=\textwidth]{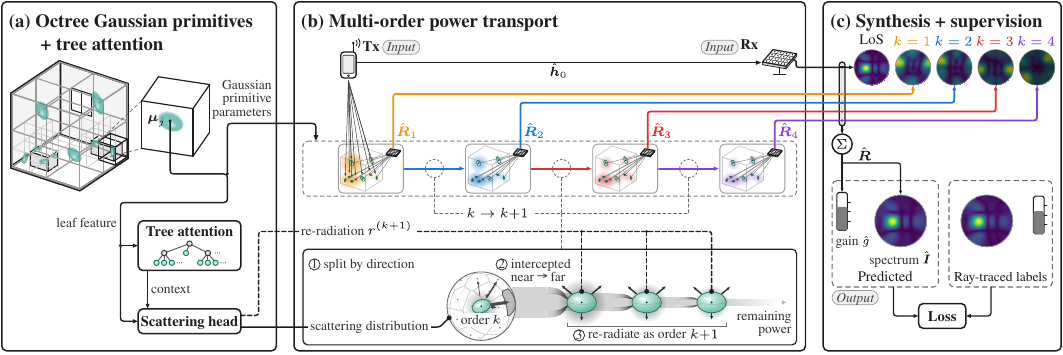}
\caption{Overview of OctCGS.
(a) Octree-anchored Gaussian primitives act as scattering nodes, and a scattering head predicts their order-wise scattering from the context of shared tree attention.
(b) Power is transported between the nodes order by order.
(c) The LoS term and the power of all orders received at the Rx position form the channel gain and the spatial spectrum, which are compared with the labels.}
\label{fig:pipeline}
\end{figure*}

All of these Gaussian splatting methods send the contribution of each primitive directly toward the Rx position,
where the primitives in between can only attenuate it or shift its phase,
and thus represent the propagation beyond the line of sight (LoS) by one wireless radiance field.
However, many paths interact with walls and other objects several times, and the last interaction sets the arrival direction at the Rx,
whereas the received power depends on every interaction and on the blockage of every path segment.
When a wall reflection precedes a cabinet reflection, the cabinet sets the arrival direction,
but the received power also depends on the wall region that the Tx illuminates.
BiWGS and RxGS represent this path by the response of a single primitive~\cite{biwgs,rxgs}, which must account for the wall-induced dependence on the Tx position,
and this dependence recurs in every primitive that represents another path via this wall.

If power is transported between primitives, the influence of the wall can be represented once for all paths that interact with it.
The response of a primitive then depends on nearby primitives, which intercept part of the power that it re-radiates,
and on large structures such as the wall, which determine the power that reaches it.
This dependence requires context from both the neighborhood of each primitive and the larger regions of the room around it.

Attention can supply such context as a learned weighted sum of the features of the other primitives, but its cost grows quadratically with the number of primitives.
BiSplat-WRF~\cite{bisplatwrf} applies attention among the points of a fixed grid of arrival directions onto which the primitive features are aggregated bilinearly,
and its attention cost is thus independent of the number of primitives.
The cells of an octree can provide this local and room-wide context:
the siblings of a leaf cover its immediate surroundings, and its ancestors cover larger regions up to the whole room.
Attention restricted to these cells has a cost linear in the number of cells, and cell splits can adapt the resolution of the octree to the scene.
Octrees organize Gaussian primitives by level of detail for rendering~\cite{octreegs}
and limit the context of attention in point cloud networks such as OctAttention~\cite{OctAttention} and OctFormer~\cite{OctFormer}.

This paper proposes OctCGS, an octree-context Gaussian splatting method with explicit multi-order propagation.
As illustrated in Fig.~\ref{fig:octree}, OctCGS anchors Gaussian primitives on an octree, and these primitives act as virtual scattering nodes.
Shared tree attention provides each node with this context, and a scattering head maps the context to the share of incoming power that each node re-radiates
and to an order-wise scattering distribution of this share over directions.
The power that has undergone exactly $k$ node-level interactions forms order $k$.
Our contributions are as follows:
\begin{itemize}
\item We propose OctCGS, which introduces explicit multi-order propagation into Gaussian splatting CKMs:
order-wise power transported between virtual scattering nodes represents multi-bounce propagation,
and the contribution of each order to the spatial spectrum and the gain can be rendered separately.
\item We develop an octree-anchored Gaussian representation whose shared tree attention gives every node local context from neighboring cells
and room-wide context from ancestor cells at a cost linear in the number of octree cells.
\item We show that OctCGS predicts spatial spectra and channel gains more accurately than BiWGS and RxGS on five indoor ray-tracing scenes,
and that the spatial-spectrum accuracy of OctCGS rises with the number of modeled orders and then saturates.
\end{itemize}

\section{Problem Formulation}
\label{sec:formulation}

We consider a static scene whose 6D CKM maps the Tx position $\boldsymbol p_{\mathrm{Tx}}\in\mathbb R^3$
and the Rx position $\boldsymbol p_{\mathrm{Rx}}\in\mathbb R^3$
to the channel knowledge of a single-input multiple-output (SIMO) link operating at a single carrier frequency with wavelength $\lambda$.
The Rx carries an array of $N$ elements with element positions $\boldsymbol\nu_n\in\mathbb R^3$ relative to the array center.
A plane wave from an arrival direction $\boldsymbol\omega$ on the unit sphere $\mathbb S^2$ around the Rx position reaches element $n$
with the phase factor $a_n(\boldsymbol\omega)=\exp(\mathrm j2\pi\boldsymbol\nu_n^{T}\boldsymbol\omega/\lambda)$,
and these factors form the array response $\boldsymbol a(\boldsymbol\omega)\in\mathbb C^{N}$.

The channel knowledge consists of the channel gain $g=10\log_{10}\|\boldsymbol h\|^2$ of the SIMO channel vector $\boldsymbol h\in\mathbb C^{N}$
and the spatial spectrum $I(\boldsymbol\omega)=|\boldsymbol a^{H}(\boldsymbol\omega)\boldsymbol h|^2$,
which is the received power over arrival directions under conventional beamforming.
The phases of all paths, including the LoS path, change with small displacements of the Tx position or the Rx position
and are modeled as independent and uniformly distributed, which gives the channel vector zero mean,
and the expected spectrum and the expected channel power follow from its covariance $\boldsymbol R=\mathbb E[\boldsymbol h\boldsymbol h^{H}]\in\mathbb C^{N\times N}$ as
\begin{equation}
\mathbb E[I(\boldsymbol\omega)]=\boldsymbol a^{H}(\boldsymbol\omega)\,\boldsymbol R\,\boldsymbol a(\boldsymbol\omega),\qquad
\mathbb E\big[\|\boldsymbol h\|^2\big]=\mathrm{tr}\,\boldsymbol R.
\label{eq:cov_quantities}
\end{equation}
The samples of $I(\boldsymbol\omega)$ on a $1^\circ$ grid of zenith and azimuth angles over the upper hemisphere form a matrix $\boldsymbol I\in\mathbb R_{+}^{90\times360}$.
The CKM is learned from training links in which both positions vary, and only the total $g$ and $\boldsymbol I$ of each link over all paths serve as labels.

\section{Proposed Method}
\label{sec:method}

OctCGS predicts the covariance $\hat{\boldsymbol R}$ of each link in three steps, and the predicted spectrum
$\hat I(\boldsymbol\omega)=\boldsymbol a^{H}(\boldsymbol\omega)\hat{\boldsymbol R}\boldsymbol a(\boldsymbol\omega)$
and gain $\hat g=10\log_{10}\mathrm{tr}\,\hat{\boldsymbol R}$ follow as in (\ref{eq:cov_quantities}).
As shown in Fig.~\ref{fig:pipeline}(a), a scattering head predicts the order-wise scattering distributions and re-radiation shares of octree-anchored Gaussian primitives,
which act as virtual scattering nodes, from the context of shared tree attention.
The power that the Tx injects into the nodes is then transported between them order by order, as illustrated in Fig.~\ref{fig:pipeline}(b),
and the LoS term and the power re-radiated toward the Rx position in all orders form $\hat{\boldsymbol R}$, as depicted in Fig.~\ref{fig:pipeline}(c).

\subsection{Octree-Anchored Gaussian Primitives}
\label{sec:octree_env}

As shown in Fig.~\ref{fig:octree}, OctCGS partitions the room bounding box by recursive splits,
each of which halves a cell along its three axes into eight sibling children.
The cells containing a given cell are its ancestors, and unsplit cells are leaves.
These nested cells range from the surroundings of a primitive to the whole room and can thus supply both the local and the room-wide context of node-level interactions.
All leaves are initialized at the same depth before the octree is refined during training, and the $M$ Gaussian primitives occupy distinct leaves.
Primitive $j$ is parameterized by the rotation and three scales of its Gaussian ellipsoid, an opacity $\alpha_j\in(0,1)$, a scattering amplitude $f_j>0$
and a center $\boldsymbol\mu_j\in\mathbb R^3$, which a learned offset from the leaf center confines to its hosting leaf $v(j)$.

\begin{figure}[!t]
\centering
\includegraphics[width=\columnwidth]{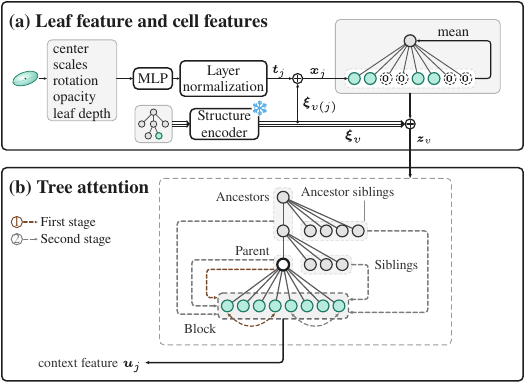}
\caption{Shared tree attention. (a) Leaf feature and cell features. (b) Context sets of the two attention stages.}
\label{fig:tree_attn}
\end{figure}

\subsection{Tree Attention for Hierarchical Context}
\label{sec:tree_attn}

Shared tree attention draws the context of each primitive from a fixed number of neighboring and ancestor cells, hence at a cost linear in the number of octree cells.
A shared multilayer perceptron (MLP) with layer normalization (LN) maps the center $\bar{\boldsymbol\mu}_j$ of primitive $j$ normalized to the room bounding box,
its scales, rotation, opacity and leaf depth to a geometry embedding $\boldsymbol t_j$.
Each octree cell $v$ carries a structure embedding $\boldsymbol\xi_v$ that encodes its position in both the octree and the room and remains untrained.
As shown in Fig.~\ref{fig:tree_attn}(a), adding the embedding $\boldsymbol\xi_{v(j)}$ of the hosting leaf $v(j)$ to $\boldsymbol t_j$ gives the leaf feature $\boldsymbol x_j=\boldsymbol t_j+\boldsymbol\xi_{v(j)}$,
which enters both the tree attention and the scattering head,
and the leaf features within each cell $v$ form its cell feature $\boldsymbol z_v$ with $\boldsymbol\xi_v$ as a positional encoding,
\begin{equation}
\boldsymbol z_v=\boldsymbol\tau_v+\boldsymbol\xi_v,\qquad
\boldsymbol\tau_v=\begin{cases}
\boldsymbol x_j, & v=v(j),\\
\boldsymbol 0, & v\ \text{empty leaf},\\
\tfrac18\sum_{v'\in\mathcal C(v)}\boldsymbol\tau_{v'}, & \text{otherwise},
\end{cases}
\label{eq:node_token}
\end{equation}
with the aggregated feature $\boldsymbol\tau_v$ and the children $\mathcal C(v)$ of $v$.

The cell features are updated in two attention stages.
The eight children $\mathcal C(o)$ of a parent cell $o$ form a block, which attends to the context set $\mathcal E=\mathcal C(o)\cup\{o\}$ in the first stage,
and each stage shares one set of parameters across all blocks.
As shown in Fig.~\ref{fig:tree_attn}(b), a fixed number of siblings of $o$, ancestors of $o$ and siblings of these ancestors are added to $\mathcal E$
in the second stage, which processes the first-stage outputs, and their bottom-up features extend the context to larger and more distant regions of the room.
Every $\boldsymbol z_v$ within the block is updated by
\begin{equation}
\boldsymbol z_v\leftarrow\boldsymbol z_v+\mathrm{LN}\Big(\mathrm{Attn}\big(\mathrm{LN}(\boldsymbol z_v),\{\mathrm{LN}(\boldsymbol z_s)\}_{s\in\mathcal E}\big)\Big),
\label{eq:attn_update}
\end{equation}
with multi-head attention $\mathrm{Attn}$.
The second-stage output for leaf $v(j)$ is the context feature $\boldsymbol u_j$ of primitive $j$
and combines the features of the cells around $v(j)$ with those of the larger regions that contain or surround it.

\subsection{Explicit Multi-Order Propagation}
\label{sec:multi_order}

Order $k$ consists of the power after $k$ node-level interactions, up to a maximum order $K$.
The scattering head is a residual MLP that maps $\boldsymbol x_j$, $\boldsymbol u_j$
and $\bar{\boldsymbol\mu}_j$ to the symmetric coefficient matrix $\boldsymbol C_j^{(k)}\in\mathbb R^{9\times9}$
of the scattering distribution of order $k=1,\dots,K$ and to a re-radiation share $r_j^{(k)}\in(0,1)$ for $k\ge2$.
The scattering distribution $P_j^{(k)}(\boldsymbol\omega,\boldsymbol\omega')$ is the density of the outgoing direction $\boldsymbol\omega'\in\mathbb S^2$
of power that reaches node $j$ from the incoming direction $\boldsymbol\omega$ after $k-1$ interactions, with both directions pointing away from the node,
\begin{align}
\psi_j^{(k)}(\boldsymbol\omega,\boldsymbol\omega')&=\boldsymbol y(\boldsymbol\omega)^{T}\boldsymbol C_j^{(k)}\boldsymbol y(\boldsymbol\omega'),\label{eq:logit}\\
P_j^{(k)}(\boldsymbol\omega,\boldsymbol\omega')&=\frac{\exp\psi_j^{(k)}(\boldsymbol\omega,\boldsymbol\omega')}
{\int_{\mathbb S^2}\exp\psi_j^{(k)}(\boldsymbol\omega,\boldsymbol\omega'')\,\mathrm d\boldsymbol\omega''},\label{eq:phase}
\end{align}
with the unnormalized log-density $\psi_j^{(k)}$ and the vector $\boldsymbol y(\cdot)\in\mathbb R^9$ of the nine real spherical harmonics up to degree two.
The scattering distributions and re-radiation shares act as learned counterparts of the reflection, diffraction and scattering laws of ray tracing.

The Tx injects the power $w_j^{(1)}=f_j^2\,\gamma_j(\boldsymbol p_{\mathrm{Tx}})$ into node $j$, where the power transfer factor $\gamma_j(\boldsymbol p)$
between a position $\boldsymbol p\in\{\boldsymbol p_{\mathrm{Tx}},\boldsymbol p_{\mathrm{Rx}}\}$
and node $j$ combines inverse-square spreading with the power transmittance of the primitives in between,
\begin{align}
\gamma_j(\boldsymbol p)&=\frac{\Theta_j(\boldsymbol p)^2}{\|\boldsymbol\mu_j-\boldsymbol p\|^2},\label{eq:edge}\\
\Theta_j(\boldsymbol p)&=\prod_{i}\big(1-\alpha_i G_i\big).\label{eq:transmittance}
\end{align}
with the amplitude transmittance $\Theta_j$ taken over the primitives $i$ on the segment between $\boldsymbol p$ and $\boldsymbol\mu_j$
and the Gaussian $G_i\in(0,1]$ of primitive $i$ projected onto the plane orthogonal to the segment and evaluated where the segment crosses this plane.

Power transport is discretized into fixed direction bins $b$ with centers $\bar{\boldsymbol\omega}_b\in\mathbb S^2$ spread nearly uniformly over the sphere.
Node $j$ emits the power $e_{j,b}^{(k)}$ into bin $b$ according to the discrete scattering distribution
$\bar P_j^{(k)}(\boldsymbol\omega,b)=\mathrm{softmax}_b\,\psi_j^{(k)}(\boldsymbol\omega,\bar{\boldsymbol\omega}_b)$,
which replaces the integral in (\ref{eq:phase}) with a summation over the bin centers,
and $e_{j,b}^{(1)}=w_j^{(1)}\bar P_j^{(1)}(\boldsymbol\omega_j^{\mathrm{Tx}},b)$
with the Tx direction $\boldsymbol\omega_j^{\mathrm{Tx}}\in\mathbb S^2$ from node $j$ to the Tx position.
The power emitted into bin $b$ is intercepted by the nodes of the interception list $\mathcal N_{j,b}$,
which holds a bounded number of nodes in the direction of bin $b$ sorted by their distance from node $j$.
Node $i\in\mathcal N_{j,b}$ intercepts the share
\begin{equation}
\beta_{j\to i}=\kappa_i\prod_{i'\prec i}\big(1-\kappa_{i'}\big),\qquad \kappa_i=1-(1-\alpha_i)^2,
\label{eq:intercept}
\end{equation}
of the emitted power, with the nodes $i'$ that precede $i$ in $\mathcal N_{j,b}$ and the power fraction $\kappa_i$ intercepted along a ray through the center of node $i$.
The incoming power $w_{i,b}^{(k)}$ of node $i$ for $k\ge2$ is the share $r_i^{(k)}$ of its power intercepted from bin $b$ in order $k-1$,
and $e_{i,b'}^{(k)}$ follows from its scattering distribution of order $k$,
\begin{align}
w_{i,b}^{(k)}&=r_i^{(k)}\sum_{j:\,i\in\mathcal N_{j,b}}\beta_{j\to i}\,e_{j,b}^{(k-1)},\label{eq:arrive}\\
e_{i,b'}^{(k)}&=\sum_{b}w_{i,b}^{(k)}\,\bar P_i^{(k)}(-\bar{\boldsymbol\omega}_b,b').\label{eq:emit}
\end{align}

\subsection{Covariance Rendering}
\label{sec:render}

The power that the nodes re-radiate toward the Rx position in order $k$ forms the covariance $\hat{\boldsymbol R}_k$ of this order,
\begin{equation}
\hat{\boldsymbol R}_k=\sum_{j=1}^{M}W_j^{(k)}\rho_j\,\boldsymbol a_j\boldsymbol a_j^{H},
\label{eq:order_cov}
\end{equation}
with the path gain $\rho_j=(\lambda/4\pi)^2\gamma_j(\boldsymbol p_{\mathrm{Rx}})$, $\boldsymbol a_j=\boldsymbol a(-\boldsymbol\omega_j^{\mathrm{Rx}})$
for the Rx direction $\boldsymbol\omega_j^{\mathrm{Rx}}\in\mathbb S^2$ from node $j$ toward the Rx position and the node weights
$W_j^{(1)}=w_j^{(1)}\,4\pi P_j^{(1)}(\boldsymbol\omega_j^{\mathrm{Tx}},\boldsymbol\omega_j^{\mathrm{Rx}})$ and
$W_j^{(k)}=\sum_b w_{j,b}^{(k)}\,4\pi P_j^{(k)}(-\bar{\boldsymbol\omega}_b,\boldsymbol\omega_j^{\mathrm{Rx}})$ for $k\ge2$,
which evaluate the scattering distribution in the Rx direction.
The primitives between the Tx position and the Rx position attenuate the LoS channel vector $\hat{\boldsymbol h}_0\in\mathbb C^N$,
\begin{equation}
\hat{\boldsymbol h}_0=\frac{\lambda}{4\pi d}\,\Theta_{\mathrm{TR}}\,\boldsymbol a_{\mathrm{LoS}},
\label{eq:los}
\end{equation}
with the Tx--Rx distance $d$, $\Theta_{\mathrm{TR}}$ of the Tx--Rx segment as in (\ref{eq:transmittance})
and the array response $\boldsymbol a_{\mathrm{LoS}}$ toward the Tx position,
whose phase differences across the elements are the only phases that enter $\hat{\boldsymbol h}_0\hat{\boldsymbol h}_0^{H}$.
Since paths with independent phases leave no cross terms in $\boldsymbol R$, the LoS term and all orders add in power,
\begin{equation}
\hat{\boldsymbol R}=\hat{\boldsymbol h}_0\hat{\boldsymbol h}_0^{H}+\sum_{k=1}^{K}\hat{\boldsymbol R}_k.
\label{eq:total_cov}
\end{equation}
Truncating (\ref{eq:total_cov}) after order $k$ gives the truncated prediction, and $\hat{\boldsymbol R}_k$ alone gives the prediction of order $k$.

\subsection{Loss and Training}
\label{sec:training}

OctCGS is trained to minimize
\begin{equation}
\mathcal L=\eta_1\|\hat{\boldsymbol I}_{\mathrm{dB}}-\boldsymbol I_{\mathrm{dB}}\|^2+\eta_2|\hat g-g|+\eta_3\big(1-\mathrm{SSIM}(\hat{\boldsymbol A},\boldsymbol A)\big),
\label{eq:loss}
\end{equation}
with the weights $\eta_1$, $\eta_2$ and $\eta_3$, the dB spectra $\hat{\boldsymbol I}_{\mathrm{dB}}$ and $\boldsymbol I_{\mathrm{dB}}$, the structural similarity (SSIM)
and the amplitude spectra $\hat{\boldsymbol A}$ and $\boldsymbol A$, which are the min--max normalized elementwise square roots of $\hat{\boldsymbol I}$ and $\boldsymbol I$.
The orders are introduced successively, each initialized from the scattering-distribution parameters of the previous order.
Primitives with large positional gradients mark inaccurately represented regions and are duplicated into empty sibling leaves
or into the children of their hosting leaves, which are split when no sibling leaf is empty.
A primitive can be moved into a child cell or the parent cell when its size is small or large relative to its cell,
which adapts the octree resolution to the primitive sizes,
and primitives with low opacity and low importance to the loss are pruned.

\section{Experiments and Results}
\label{sec:experiments}

\subsection{Datasets and Setup}
\label{sec:setup}

A multi-scene dataset of five indoor scenes was generated with Sionna RT~\cite{sionna} at 6~GHz,
and its channel vectors contain the LoS path and the paths with at most six specular reflections.
The Tx carries a single antenna, and the Rx carries a horizontal $4\times4$ array with half-wavelength spacing.
Each scene was split into 14,400 training links, 2,400 validation links and 800 test links.

Two public datasets complement the multi-scene dataset.
The conference-room dataset of RxGS~\cite{rxgs} was simulated with Sionna RT at 2.4~GHz for an $8\times8$ Rx array with random element errors
and contains spatial spectrum images for 21 Rx positions and 5,089 Tx positions, of which 4,071 form the training split.
The BLE dataset of NeRF$^2$~\cite{nerf2} records the received signal strength indicator (RSSI) of 6,000 Tx positions at 21 Rx gateways,
and its official test split comprises 8,400 detected Tx--gateway pairs.

Accuracy on the multi-scene dataset is measured by the SSIM between $\hat{\boldsymbol A}$ and $\boldsymbol A$ and by the gain mean absolute error (MAE) in dB.
We reimplemented BiWGS~\cite{biwgs} following its paper and built RxGS~\cite{rxgs} on its official code with the GSRF~\cite{yang2025gsrf} backbone,
and we retrained both adapted baselines on the multi-scene dataset.
We set $K=4$ for OctCGS according to the results of Section~\ref{sec:order_effect}.

\subsection{Multi-Scene Comparison}
\label{sec:multiscene}

As shown in Table~\ref{tab:main}, OctCGS attains the best SSIM and gain MAE in every scene,
with an SSIM 0.13 to 0.23 higher and a gain MAE 0.86 to 3.79 dB lower than the best baseline per scene and metric.

\begin{table}[!t]
\centering
\caption{Results on the five scenes}
\label{tab:main}
\setlength{\tabcolsep}{2pt}
\scriptsize
\begin{tabular}{llcc}
\toprule
Scene & Method & SSIM $\uparrow$ & Gain MAE (dB) $\downarrow$ \\
\midrule
Parallel-wall & BiWGS & 0.554 & 3.23 \\
 & RxGS & 0.458 & 3.48 \\
 & OctCGS & \textbf{0.714} & \textbf{1.80} \\
\midrule
Angled-wall & BiWGS & 0.530 & 4.02 \\
 & RxGS & 0.435 & 3.72 \\
 & OctCGS & \textbf{0.716} & \textbf{1.77} \\
\midrule
T-shaped & BiWGS & 0.614 & 2.39 \\
 & RxGS & 0.451 & 2.95 \\
 & OctCGS & \textbf{0.762} & \textbf{1.53} \\
\midrule
Cluttered & BiWGS & 0.539 & 4.04 \\
 & RxGS & 0.417 & 5.03 \\
 & OctCGS & \textbf{0.670} & \textbf{2.42} \\
\midrule
Six-room & BiWGS & 0.330 & 8.63 \\
 & RxGS & 0.340 & 9.62 \\
 & OctCGS & \textbf{0.567} & \textbf{4.84} \\
\bottomrule
\end{tabular}

\end{table}

\begin{figure}[!t]
\centering
\includegraphics[width=\columnwidth]{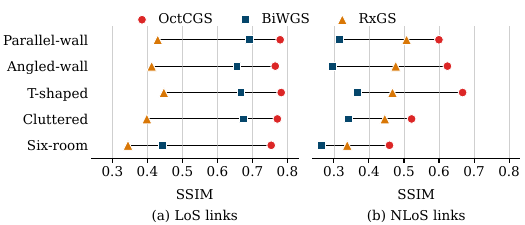}
\caption{SSIM on LoS links (a) and NLoS links (b) in the five scenes}
\label{fig:link_types}
\end{figure}

As shown in Fig.~\ref{fig:link_types}, the SSIM of OctCGS is also the highest on both LoS and non-LoS (NLoS) links in every scene.
The margin of OctCGS over RxGS is larger on LoS links than on NLoS links in all five scenes,
whereas the margin over BiWGS is larger on NLoS links in four scenes.

\begin{figure}[!t]
\centering
\includegraphics[width=\columnwidth]{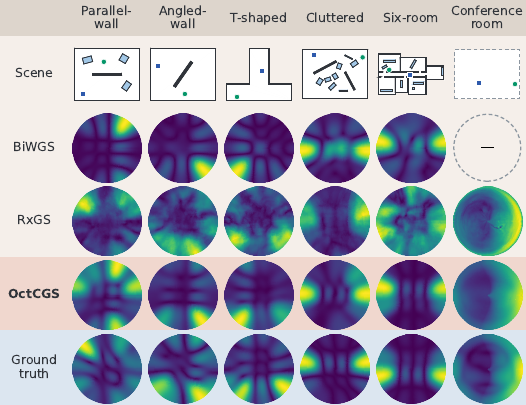}
\caption{Spatial spectra of a randomly drawn NLoS link in each scene and of one conference-room sample.
The top row marks the Tx (circle), the Rx (square) and the conference-room Tx region (dashed).}
\label{fig:spectrum_examples}
\end{figure}

As shown in Fig.~\ref{fig:spectrum_examples}, OctCGS reproduces the strongest lobes of the ground truth in four scenes,
whereas the RxGS spectra contain irregular small-scale fluctuations and three BiWGS spectra show regular stripes.

\subsection{Public Datasets}
\label{sec:benchmarks}

OctCGS accounts for the random element errors of the conference-room array by replacing $\boldsymbol\nu_n$ with 64 learned virtual element positions
and $\boldsymbol a(\boldsymbol\omega)$ in (\ref{eq:cov_quantities}) with a separate learned complex vector in $\mathbb C^{64}$ for each spectrum pixel.
The peak signal-to-noise ratio (PSNR) and the SSIM of OctCGS in Table~\ref{tab:rxgs} reach 24.95 dB and 0.850,
compared to 22.34 dB and 0.776 for RxGS (GSRF).
The conference-room values of the methods marked $\dagger$, among them WRF-GS+~\cite{wrfgsplus} and RF-3DGS~\cite{rf3dgs}, are reproduced from the RxGS paper~\cite{rxgs}.

\begin{table}[!t]
\centering
\caption{Results on the conference-room and BLE datasets}
\label{tab:rxgs}
\setlength{\tabcolsep}{3pt}
\scriptsize
\begin{tabular}{lccccc}
\toprule
 & & \multicolumn{3}{c}{Conference room} & BLE RSSI \\
\cmidrule(lr){3-5}\cmidrule(lr){6-6}
Method & \#Models & PSNR & MSE & SSIM & MAE \\
 & & (dB) $\uparrow$ & ($\times 10^{-3}$) $\downarrow$ & $\uparrow$ & (dB) $\downarrow$ \\
\midrule
GSRF$^\dagger$ & 21 & 23.29 & 5.2 & 0.831 & 3.89 \\
RF-3DGS$^\dagger$ & 21 & 23.11 & 5.4 & 0.801 & -- \\
WRF-GS+$^\dagger$ & 21 & \textbf{25.81} & \textbf{3.4} & \textbf{0.892} & 3.48 \\
SwiftWRF & 21 & -- & -- & -- & \textbf{3.29} \\
\midrule
NeRF$^2$$^\dagger$ & 1 & 24.58 & 4.0 & \underline{0.868} & -- \\
RxGS (RF-3DGS)$^\dagger$ & 1 & 22.22 & 6.8 & 0.791 & -- \\
RxGS (WRF-GS+)$^\dagger$ & 1 & 24.27 & 4.8 & 0.860 & -- \\
RxGS (GSRF) & 1 & 22.34 & 6.68 & 0.776 & 3.69 \\
OctCGS & 1 & \underline{24.95} & \underline{3.95} & 0.850 & \underline{3.43} \\
\bottomrule
\end{tabular}

\end{table}

The RSSI serves as the label instead of $g$ on the BLE dataset.
WRF-GS+, GSRF and SwiftWRF~\cite{swiftwrf} assume a single Rx position and were hence trained separately for each gateway on the official split.
The RSSI MAE of OctCGS in Table~\ref{tab:rxgs} is 3.43 dB, lower than the 3.69 dB of RxGS (GSRF).

\subsection{Effect of the Maximum Number of Node-Level Interactions}
\label{sec:order_effect}

\begin{figure}[!t]
\centering
\includegraphics[width=\columnwidth]{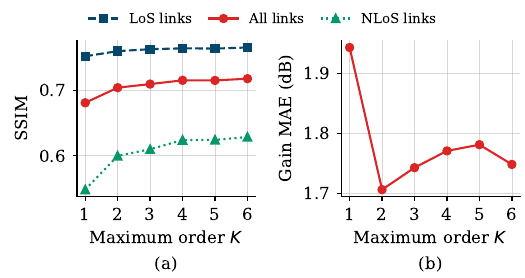}
\caption{SSIM (a) and gain MAE (b) versus the maximum order $K$}
\label{fig:order_sweep}
\end{figure}

Separate models with $K=1,\dots,6$ were trained on the Angled-wall scene to measure the effect of additional node-level interactions.
Fig.~\ref{fig:order_sweep} shows that the SSIM rises from 0.681 at $K=1$ to 0.716 at $K=4$ and remains nearly constant for larger $K$.
The improvement occurs mainly on NLoS links, whose SSIM increases from 0.548 at $K=1$ to 0.624 at $K=4$, whereas the SSIM of LoS links grows only from 0.753 to 0.765.
The gain MAE falls from 1.94 dB at $K=1$ to 1.71 dB at $K=2$ and then varies by less than 0.08 dB.
We therefore chose $K=4$, where the SSIM saturates.

\begin{figure}[!t]
\centering
\includegraphics[width=\columnwidth]{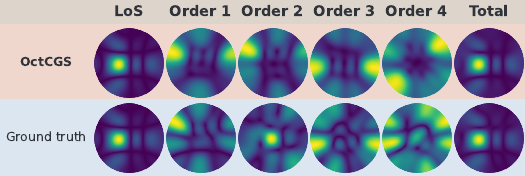}
\caption{Peak-normalized order-wise spectra of OctCGS and the ray-traced ground truth for the link with the highest SSIM}
\label{fig:order_components}
\end{figure}

The prediction truncated after order 2 in Table~\ref{tab:order_energy} already attains an SSIM of 0.712
and a 1.99 dB gain MAE, compared to 0.716 and 1.77 dB with all four orders.
The energy share of order $k$ is the link-averaged ratio of its covariance trace to the total covariance trace of the LoS term and all orders
(four node-level orders for OctCGS and six reflection orders for ray tracing).
The energy shares of OctCGS for the LoS term and orders 1 to 3 deviate from those of ray tracing by at most 2.4 percentage points.
As shown in Fig.~\ref{fig:order_components}, several order-wise lobes of OctCGS coincide with those of ray tracing,
and the total OctCGS spectrum reproduces the strongest lobes of the total ray-tracing spectrum.

\begin{table}[!t]
\centering
\caption{Order truncation and energy share of the model with $K=4$ on the Angled-wall scene}
\label{tab:order_energy}
\setlength{\tabcolsep}{2pt}
\scriptsize
\begin{tabular}{lcccc}
\toprule
 & \multicolumn{2}{c}{LoS and orders $\leq k$} & \multicolumn{2}{c}{Energy share of order $k$ (\%)} \\
\cmidrule(lr){2-3}\cmidrule(lr){4-5}
Order $k$ & SSIM & Gain MAE (dB) & OctCGS & Ray tracing \\
\midrule
0 (LoS) & 0.553 & 24.39 & 47.3 & 49.5 \\
1 & 0.695 & 3.44 & 25.9 & 28.3 \\
2 & 0.712 & 1.99 & 15.1 & 15.3 \\
3 & 0.715 & 1.85 & 6.0 & 5.1 \\
4 & 0.716 & 1.77 & 5.7 & 1.3 \\
\bottomrule
\end{tabular}

\end{table}

\section{Conclusion}
\label{sec:conclusion}

OctCGS constructs 6D CKMs by order-wise power transport between Gaussian primitives anchored on an octree,
whose neighboring and ancestor cells give every primitive local and room-wide context through shared tree attention.
Its spatial spectra and channel gains are more accurate than those of the baselines in five indoor ray-tracing scenes.
The higher orders of OctCGS improve the predicted spatial spectra mainly on NLoS links,
and its node-level interactions allow a Gaussian splatting CKM to represent multi-bounce propagation order by order.

\bibliographystyle{IEEEtran}
\bibliography{refs}

\end{document}